\documentclass[sigconf]{acmart}

\usepackage{color}
\usepackage[linesnumbered]{algorithm2e}
\usepackage[caption=false]{subfig}
\usepackage{graphicx}
\usepackage{amsfonts}
 \usepackage{multirow}
 \usepackage{pgfplots}
 \usepackage{color}
 \usepgfplotslibrary{external} 
\copyrightyear{2019}
\acmYear{2019} 
\setcopyright{iw3c2w3}
\acmConference[WWW '19]{Proceedings of the 2019 World Wide Web Conference}{May 13--17, 2019}{San Francisco, CA, USA}
\acmBooktitle{Proceedings of the 2019 World Wide Web Conference (WWW '19), May 13--17, 2019, San Francisco, CA, USA}
\acmPrice{}
\acmDOI{10.1145/3308558.3313641}
\acmISBN{978-1-4503-6674-8/19/05}

\title{The Music Streaming Sessions Dataset}

  \author{Brian Brost} 
  \affiliation{Spotify Research, London} 
  \email{brianbrost@spotify.com}
  
  \author{Rishabh Mehrotra} 
  \affiliation{Spotify Research, London} 
  \email{rishabhm@spotify.com}
  
  \author{Tristan Jehan} 
  \affiliation{Spotify Research, NY} 
  \email{tjehan@spotify.com}

\begin{document}

\begin{abstract}
%!TEX root = ./Paper-NIPS.tex

At the core of many important machine learning problems faced by online streaming services is a need to model how users interact with the content they are served.
Unfortunately, there are no public datasets currently available that enable researchers to explore this topic. In order to spur that research, we release the Music Streaming Sessions Dataset (MSSD), which consists of 160 million listening sessions and associated user actions. Furthermore, we provide audio features and metadata for the approximately 3.7 million unique tracks referred to in the logs. This is the largest collection of such track metadata currently available to the public. This dataset enables research on important problems including how to model user listening and interaction behaviour in streaming, as well as Music Information Retrieval (MIR), and session-based sequential recommendations. Additionally, a subset of sessions were collected using a uniformly random recommendation setting, enabling their use for counterfactual evaluation of such sequential recommendations. Finally, we provide an analysis of user behavior and suggest further research problems which can be addressed using the dataset.

\end{abstract}
\maketitle

\section{Introduction}
\label{sec:Introduction}

%!TEX root = ./Paper-NIPS.tex

A long-standing and central challenge for online services is to understand and model how users behave \cite{webb2001machine}. For web search and online advertising, this has led to a large body of work on click modeling, most of which would not have been possible without access to publicly available click logs \cite{chuklin2015click}. Modeling user behaviour is similarly important to streaming services \cite{yu2006understanding,vigliensoni2017music}, but to the best of our knowledge there are no user interaction datasets currently available to the public. This is particularly limiting when it comes to designing recommender systems, where the use of implicit feedback is often critical \cite{oard1998implicit}.

Motivated by the paucity of user interaction logs in streaming, we release the Music Streaming Sessions Dataset (MSSD), which consists of over 160 million listening sessions with associated user interaction information. In addition, we provide audio features and metadata for the approximately 3.7 million unique tracks referred to in the logs, making this the largest collection of such track metadata currently available to the public.  A useful characteristic of the dataset is that a subset of the sessions contained in the log were obtained using a uniformly random shuffle, enabling research on counterfactual methods. Lastly, we provide snapshots of the playlists from which this subset of sessions was streamed.

By providing the sequences of track plays, the smaller pools of tracks which were available for recommendation during the session, and the probabilities each of those tracks had of being played next, we allow this dataset to be used as a test-bed for sequential recommendation research, and in particular with counterfactual approaches to the problem. The rise of counterfactual methods \cite{bottou2013counterfactual} dramatically improved the speed of experimentation for online services, however there is still very little data available to academic researchers in that area, and no data at all for the specific problem of sequential recommendation. 

One of the main ambitions of this dataset release is to enable public research on two central challenges facing music streaming services, namely understanding when a user will 1) skip a track, and 2) move from one listening context to another. Related to this dataset we therefore also organized a skip prediction machine learning challenge for the 2019 WSDM Cup. More generally, we expect that this dataset will spur further research in the area of Music Information Retrieval (MIR), where it will function as a partial expansion of the Million Song Dataset (MSD) \cite{bertin2011million}. For the first time we provide a dataset that links track audio features with user listening behaviour. Finally, sequentially recommending items for users is of particular importance to music streaming services but also to other types of services, such as news or e-commerce, for example in next-basket recommendation. We believe the underlying solutions for these cases would be comparable to those for music.

Our contributions are therefore threefold: (1) we provide the only dataset of streaming logs and interactions currently available to the public in Section \ref{sec:Dataset}; (2) we provide an analysis of the logs in Section \ref{sec:Analysis}; and (3) we identify important research questions that can be addressed using this dataset in Section \ref{sec:Problems}. 

\section{Related Work}
\label{sec:RelatedWork}

%!TEX root = ./Paper-NIPS.tex

Click log releases have played a key role in allowing the development of sophisticated click models for web search and advertising applications, with companies including Microsoft, Yahoo, Yandex, and Criteo all providing such logs to the academic research community. An overview of the click modeling literature, and available datasets is provided in \cite{chuklin2015click}. 

In the context of recommender systems there are fewer such interaction logs available. Instead, recommender systems datasets have tended to contain explicit ratings \cite{bell2007lessons,harper2016movielens,dror2012yahoo}. Some streaming logs containing user interactions have been released, for example the XING dataset which was part of the 2016 Recsys Challenge \cite{abel2016recsys}. However, this dataset was in the field of job recommendations and is no longer publicly available. 

There are several music listening histories datasets, most of which were crawled from from social media such as last.fm or twitter, however none of these datasets contain user interactions beyond what songs were listened to \cite{celma2010music,schedl2013leveraging,hauger2013million,schedl2016lfm,vigliensoni2017music}. Similarly, the kkbox dataset from the 2018 WSDM Cup challenge provides listening logs \cite{chen2018wsdm}, but these logs are not timestamped, nor do they provide user interactions beyond what tracks were listened to. Thus, although there are some music listening logs, crucially none of these contain information about how users interacted with the tracks they listened to, showing instead only what tracks users were exposed to. 

An important feature of our dataset is that a subset of the sessions contained in the logs were collected from users experiencing a uniformly random shuffle function. This allows our dataset to be used in connection with so-called counterfactual methods. These have the potential to allow offline evaluations of new algorithms while only requiring some randomization of the deployed algorithm's output \cite{bottou2013counterfactual}. If the cost of randomizing the deployed algorithm's output is not too great, this can provide a practical way of obtaining results that are potentially as reliable as those obtained from A/B testing, but without requiring the variants being tested to be production ready \cite{gilotte2018offline}. This can allow tests of new algorithms to be carried out more quickly and efficiently than what is currently possible. To the best of our knowledge there are no other public datasets except the counterfactual test-bed from \cite{lefortier2016large}. Our dataset distinguishes itself since it is focused on the novel problem of applying counterfactual methods to sequential recommendations. Counterfactual evaluation of sequenced recommendations provides an interesting challenge since a naive approach would suffer from a growth in the propensities exponential in the length of the sequence.

\section{The Music Streaming Sessions Dataset}
\label{sec:Dataset}

%!TEX root = ./Paper-NIPS.tex

The MSSD consists of 160 million streaming sessions with associated user interactions, audio features and metadata describing the tracks streamed during the sessions, and snapshots of the playlists listened to during the sessions. The dataset is hosted at \href{http://research.spotify.com/datasets/music-streaming-sessions}{http://research.spotify.com/datasets/music-streaming-sessions} together with a set of tools for working with the dataset.

The streaming sessions are stored in a log, where each row of the log contains a session id, a timestamp, contextual information about the stream, the track and context id's, and the timing and type of user interactions within the stream. A schema for the log is provided in Table \ref{tab:log_schema}. Each session is defined to be a period of listening with no more than 60 seconds of inactivity between consecutive tracks. Additionally for this dataset we set a cut off of at most 20 tracks per session as part of our privacy strategy. Sessions included in the dataset are sampled uniformly at random from eligible listening sessions on the contexts included in the dataset over an 8 week period. We exclude sessions that include tracks which did not meet a minimum popularity threshold. The sessions in this dataset are sampled from radio, personalized recommendation mixes, the user's own collections, and 100 of the most popular playlists on a major music streaming service. The logs therefore contain a mix of listening sessions based on user's personally curated collections; expertly curated playlists; contextual, but non-personalized recommendations; and finally, personalized recommendations. 

\begin{table*}[t]
\centering
\resizebox{\linewidth}{!} 
{
\begin{tabular}{ | l | l | r | } \hline
  Column name & Column description & Example value \\ \hline
  session id & unique session identifier & 57 \textunderscore 55129e3f-29bf-4ef6-aa72-d140333eac9c
 \\
  session position & position of track within session & 18 \\
  session length & length of session & 20 \\
  track id & unique track identifier & t \textunderscore aae12819-de17-4dd3-97b0-cad4dd7b9a56
 \\
  skip 1 & whether the track was only played very briefly & false\\
  skip 2 &whether the track was only played briefly & false\\
  skip 3 & whether most of the track was played & true \\
  not skipped & whether the track was played in its entirety & false \\
  context switch & whether the user changed context between the previous row and the current row & true \\
  no pause & whether there was no pause between playback of the previous track and current track & false \\
  short pause & whether there was a short pause between playback of the previous track and current track & true \\
  long pause & whether there was a long pause between playback of the previous track and current track & true \\
  num seekfwd & the number of times the user scrubbed forward during playback & 0 \\
  num seekbk &  the number of times the user scrubbed backward during playback & 3 \\
  shuffle & whether the track was played with shuffle mode activated & false \\
  hour of day & hour of day (integers between 0 and 23) & 18 \\
  date & date in YYYY-MM-DD format & 2018-09-10 \\
  premium & whether the user was on premium or not & true \\
  context type & what type of context the playback occurred within & catalog \\
  reason start & cause of this track play starting & forward button \\
  reason end & cause of this track play ending & track done \\
  uniform random & whether shuffle would be uniformly random for this session & false \\ \hline
\end{tabular}
}
\caption{Schema for the interaction log}
\label{tab:log_schema}
\end{table*}

For each track contained in the sessions, we provide audio features and metadata describing the track. This part of our dataset can be regarded as a partial update and expansion of the Million Song Dataset (MSD) \cite{bertin2011million}. We provide approximately 3.7 million tracks, and many new features not included in the Million Song Dataset. These include features like acousticness, a measure of confidence that the track is acoustic; downbeats, estimated timestamps for the downbeats in a track; and valence, a measure of how positive a track sounds. The schema for the track features is provided in Table \ref{tab:track_schema}, features which did not exist in the Million Song Dataset are bolded. More detailed descriptions of the features are available in \cite{bertin2011million} and on our dataset website. 

\begin{table}[t]
\centering
{
\begin{tabular}{ | c | c | c | } \hline
  track id & duration & release year \\
  popularity & \textbf{acousticness} & \textbf{beat strength} \\
  \textbf{bounciness} & danceability & \textbf{dyn range mean} \\
  energy & \textbf{flatness} & \textbf{instrumentalness} \\
  key & \textbf{liveness} & loudness \\
  \textbf{mechanism} & mode & \textbf{organism} \\
  \textbf{speechiness} & tempo & time signature \\
  valence & \textbf{acoustic vector} & \\ \hline
\end{tabular}
}
\caption{Features for the track metadata}
\label{tab:track_schema}
\end{table}

\subsection{Uniform Random shuffle subset}
\label{ss:counterfactual}

As noted earlier, counterfactual methods have the potential to allow offline evaluations of new algorithms if the deployed algorithm's output is randomized and the propensities are known \cite{bottou2013counterfactual}. The standard shuffle function used during streaming on our service is not uniformly random, however we have collected and labelled a subset of the sessions contained in this log using a uniformly random shuffle. 

With this dataset we focus on the novel problem of applying counterfactual methods to sequential recommendations. Shuffled listening sessions provide a nice means of collecting randomized observations, without seriously harming the user's experience since the user already expects a somewhat random experience. However good and bad sequences of tracks can still be distinguished by a variety of session level and track play level metrics such as session length and skip rate. 

For each session using the uniformly random shuffle function we include snapshots of the tracks contained in the playlist and their positions in the playlists. This allows us to use the logs and playlist snapshots for counterfactual evaluations of sequential recommendations as outlined in Section \ref{ss:counterfactual}. %Table \ref{tab:playlist_schema} outlines the information contained in the snapshots.

\section{Dataset Analysis}
\label{sec:Analysis}

%!TEX root = ./Paper-NIPS.tex

Here we provide summary statistics and figures for the dataset, highlight some of its properties, and explore implications for future user modeling efforts. The logs detail how users behave in terms of two particularly important forms of implicit feedback: user skip behavior, and their context switch behavior, i.e. when users change from one listening context or playlist to another. 

\subsection{Summary Statistics}

\begin{figure}
            \includegraphics[width=\columnwidth]{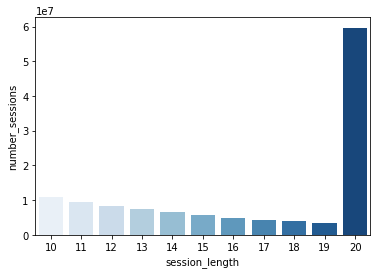}
            \caption
            {Distribution of session lengths}    
            \label{fig:sessionlengths}
\end{figure}

The dataset consists of 160 million sessions with lengths varying between 10 to 20 interactions. Recall that a session is defined as a period of listening interrupted by no more than 60 seconds between consecutive playbacks. Sessions shorter than this limit are excluded to try to increase the amount of information available early in sessions to predict a user's interactions later in the session. Recall that longer sessions are excluded as part of our privacy strategy. Figure \ref{fig:sessionlengths} shows that the proportion of sessions of a given length decreases steadily with increasing length, although sessions of length 20 are much more common since sessions that were longer are capped at this length. 

\subsection{User Interactions}

We begin by exploring user skip behavior. In this dataset we provide a variety of different skip definitions, for this analysis we use the skip \textunderscore 1 threshold. The distribution of skip rates across tracks is given in Figure \ref{fig:lengthskip}.

\begin{figure}
            \includegraphics[width=\columnwidth]{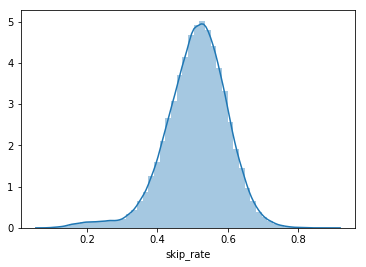}
            \caption
            {Distribution of skip rate across tracks}    
            \label{fig:lengthskip}
\end{figure}

In general, skip rates are higher for longer sessions, as shown in Figure \ref{fig:lengthskip}. 
However, whereas skip rates are higher for longer sessions, skip rates are not substantially higher earlier in sessions, remaining relatively constant at a non-skip rate between 34\% and 35\% for all session positions.

\begin{figure}
            \includegraphics[width=\columnwidth]{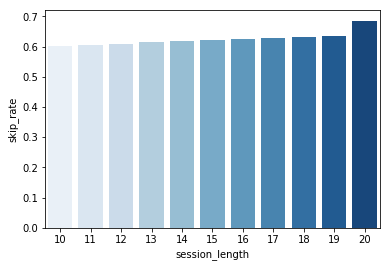}
            \caption
            {Session length against skip rate}    
            \label{fig:lengthskip}
\end{figure}

\begin{figure}
            \includegraphics[width=\columnwidth]{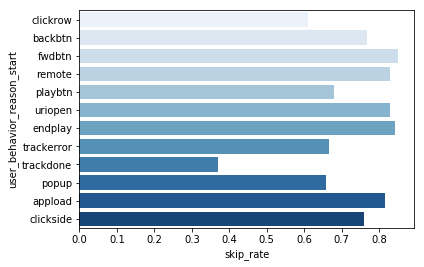}
            \caption
            {Skip rate against reason for start of playback}    
            \label{fig:reasonstartskip}
\end{figure}

Whereas it might be expected that a user explicitly selecting a song would lead to a low probability of that song being skipped, we see from Figure \ref{fig:reasonstartskip} that the reason for a track starting that is most closely associated with a track not being skipped is that the previous track was listened to completion. This may be partly explained by the fact that although a user who explicitly selected a song wanted to listen to that track, a user who is in a so-called lean-back listening mode, is even less likely to skip. A deeper analysis of such listening modes is an interesting challenge for MIR researchers. 

\begin{figure}
            \includegraphics[width=\columnwidth]{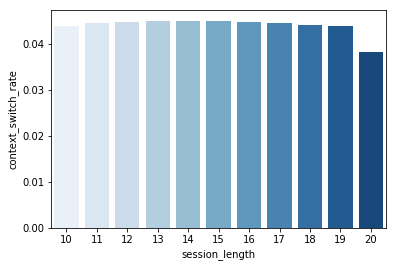}
            \caption
            {Session length against context switch rate}    
            \label{fig:sessionlengthcontextrate}
\end{figure}

\begin{figure}
            \includegraphics[width=\columnwidth]{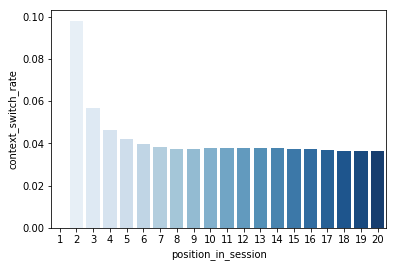}
            \caption
            {Session position against context switch rate}    
            \label{fig:sessionposcontextrate}
\end{figure}

\begin{figure}
            \includegraphics[width=\columnwidth]{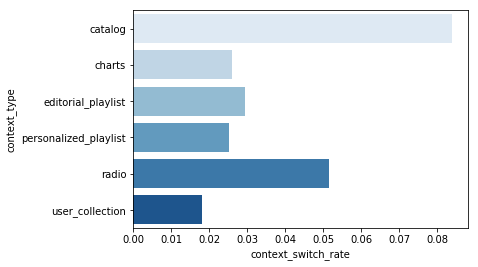}
            \caption
            {Context switch rate against source context}    
            \label{fig:sourcecontextcontextrate}
\end{figure}

Whereas skip rates are highly correlated with session length, this is less obviously the case for the context switch rate, as shown in Figure \ref{fig:sessionlengthcontextrate}, with the exception of sessions of length 20, which due to the cutoff include potentially longer sessions. Here we see a substantially lower context switch rate, possibly reflecting extreme lean-back sessions, such as sleep playlists. Despite this, we see in Figure \ref{fig:sessionposcontextrate} that context switches are much more likely at the beginning of a session. Finally, we see in Figure \ref{fig:sourcecontextcontextrate} that the likelihood of a user switching context depends heavily on the source context. Understanding the causes and directions of users context switches is an important challenge for music streaming services. 

\subsection{Track Features}

\begin{figure*}
            \includegraphics[width=\linewidth]{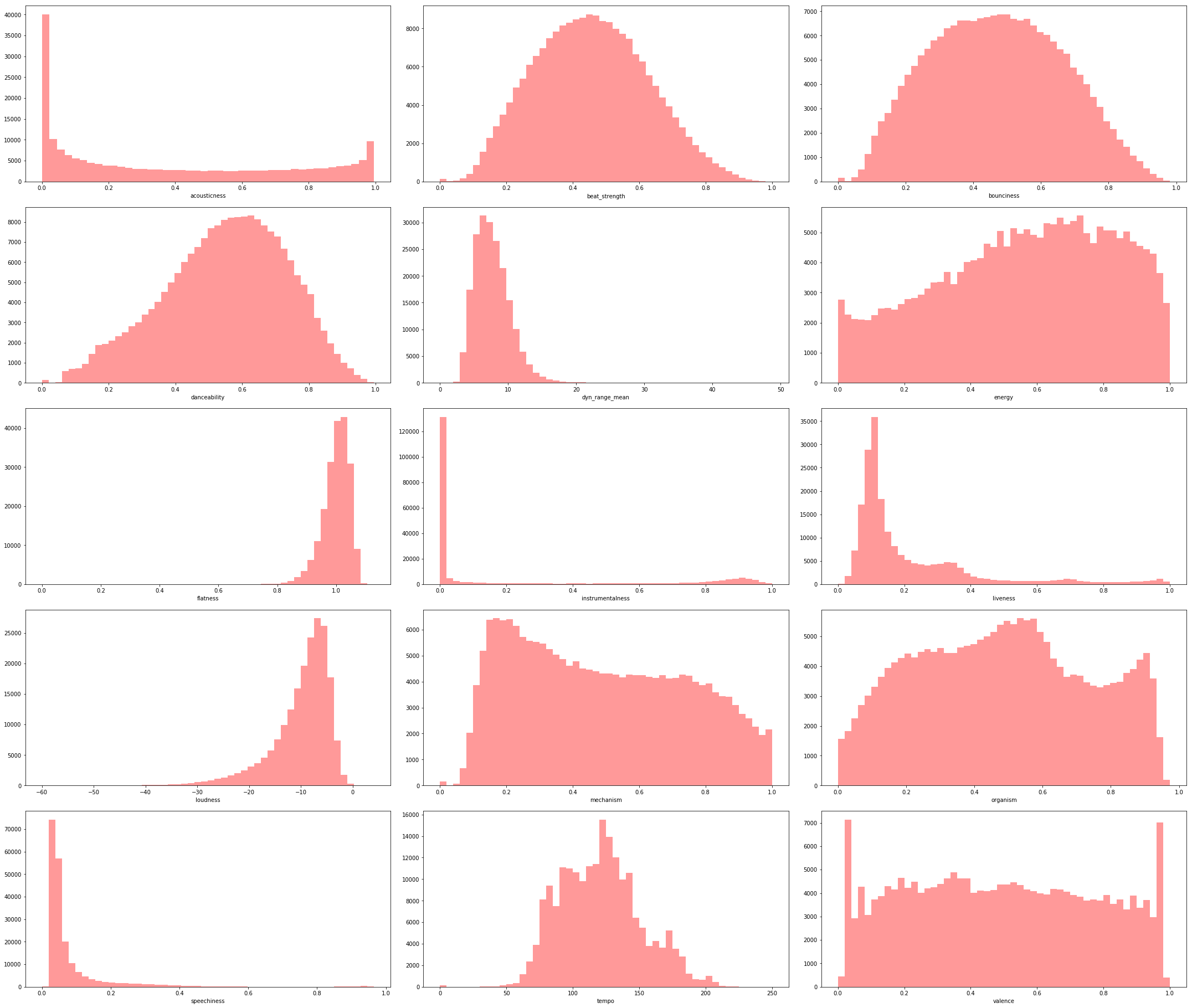}
            \caption
            {Distributions of feature values across tracks for selected audio features}    
            \label{fig:featurevaluedistributions}
\end{figure*}

Figure \ref{fig:featurevaluedistributions} shows the distributions of feature values across tracks for selected audio features. The distributions for the remaining audio features are included on the dataset website. Features display dramatically different distributions, with a feature like instrumentalness bimodal and highly skewed, with most tracks having a very low value indicating an extremely low probability estimate of the track being an instrumental track, and relatively few tracks displaying a high estimated probability of being instrumental.

Such a large number of audio features and the variations in their distribution could inspire future research on understanding how user consumption is linked with content level features. While most user engagement studies in the past have looked at patterns across recommended items, jointly performing such analysis on content features would make for an interesting insightful discussion.

%\section{Leveraging \& Predicting User Interactions}
%\label{sec:UserModels}

%\input{./04-UserModeling}

\section{Further Research Problems}
\label{sec:Problems}

%!TEX root = ./Paper-NIPS.tex

While we have highlighted issues around user interactions with a sequential recommender, specifically predicting user interactions (skips) and context switches, the released dataset allows us to go beyond these problems, and can serve as a benchmark dataset for researchers interested in various other research areas. Here we list a few key potential research areas.\\

\noindent\textbf{1. Session based Sequential Recommendations}\\
While most machine learning techniques focus on either item recommendation or even set recommendations aimed at increasing instantaneous user satisfaction, we believe that this dataset will enable researchers to propose novel techniques aimed at optimizing user sessions, and more long term satisfaction.\\

\noindent\textbf{2. User Intervention in Automated Systems}\\ For each session, the dataset provides user interaction with each track, which opens up interesting problems around human interventions with automated systems. Indeed, the user might begin by relying on the recommender system to suggest content, but at some point decide to intervene, either slightly (e.g. only skip a track) or to a larger extent (e.g. stop using the recommender system and move to their own playlist). As we develop advanced automated systems, we need to understand user reactions to them and comprehend, predict and leverage user interventions, in an attempt to not only reduce the need for further interventions but also to optimize the system more efficiently.\\

\noindent\textbf{3. Offline Evaluation of Recommender Systems}\\ Metrics and reward models of machine learning systems relies heavily on user feedback. The detailed per item meta-data allows researchers to develop advanced evaluation metrics and reward functions for systems to learn from.\\

\noindent\textbf{4. User Journeys}\\
The session level data enables research on carefully constructing user journeys not only in the general space of recommendation items, but also in the space of user moods. Indeed, research on music therapy has highlighted the impact of positive music \cite{center2005music}, and this dataset enables researchers to develop carefully crafted user journeys to affect user moods.\\

\noindent\textbf{5. Proactive Recommendations}\\
Anticipating user actions and interventions allows machine learning systems to be proactive in recommendations. So far, research on proactive recommendations has been conducted in the industry by researchers having access to large scale user interaction logs. We hope the dataset enables academic researchers to conduct research on proactive systems.\\

\noindent\textbf{6. Counterfactual Evaluation}\\
Most major web companies rely on large scale A/B tests for model development and iteration. Counterfactual estimation of metrics has recently enabled researchers to make offline predictions of online metrics.  We advocate the use of counterfactual estimation techniques for unbiased offline evaluation of systems using our dataset. Compared to A/B tests, offline evaluation allows multiple models to be evaluated on the same log, without the need to be run online. Effectively, counterfactual estimation techniques make it possible to run many A/B tests simultaneously, leading to substantial increase in experimentation agility. Developing good ounterfactual estimators for sequential recommendations is an open research problem for which this dataset can function as a test bed.

Owing to the heterogeneous data released as part of this dataset (user sessions, audio featres, user contexts, etc), we envision myriad future use-cases of this dataset, beyond the six broader areas highlighted above.

%\section{Session Modelling} 
%\label{sec:SessionModelling}

%\input{./05-SessionModelling}

%\section{Experimental Evaluation}
%\label{sec:ExperimentalEvaluation}

%\input{./06-Experiments}

\section{Conclusion}
\label{sec:Conclusion}

%!TEX root = ./Paper-NIPS.tex

The problems of understanding, modeling and predicting how users interact with content on streaming services has until now been understudied, mainly because of a lack of access to data. With the paper, we provide the only dataset of streaming logs and user interactions currently available to the research community. 

With this dataset we address two key problems facing streaming services, namely how to predict user skips and context switches. We provide an analysis of the dataset in terms of these problems. In particular, we illustrated that skip behavior in a session is correlated with prior skip behavior within that session. We believe that a thorough investigation of these forms of user behavior is an important future research challenge. 

Finally, we identify other important research questions that can be addressed using the dataset. We believe that the task of providing session-based sequential recommendations is of particular importance, due to the need for fast learning from limited user-based information when onboarding new users. 

While the dataset enables new types research directions, compromises were made for privacy and commercial reasons. An interesting future extension of this dataset would be to provide more precision on the skip information than the currently bucketed times. That level of information could allow to predict moments in the track where skips are most likely to occur, which could be of great value for generative and interactive music models.

\newpage

\small
\bibliographystyle{plain} 
\bibliography{Bib} 

\end{document}